\documentclass[aps,twocolumn,epsf,floats,pre,superscriptaddress,nofootinbib]{revtex4-1}

\usepackage{graphicx}
\usepackage{amsmath}
\usepackage{microtype}
\usepackage{tikz}
\usepackage[compat=1.1.0]{tikz-feynman}
\usepackage{commath}
\usepackage{enumitem}
\usepackage{amsfonts}
\usepackage{hyperref}
\usepackage{mathtools}
\usepackage{scrtime}
\usepackage[dont-mess-around]{fnpct}
\usepackage{import}
\usepackage{float}
\usepackage{etoolbox}
\usepackage{placeins}

\newcommand{\vx}{\boldsymbol{x}}

\newcommand{\bnabla}{\boldsymbol{\nabla}}

\begin{document}

\title{
From noise on the sites to noise on the links: discretizing the conserved Kardar-Parisi-Zhang equation in real space}

\author{Andrea Cavagna}
\affiliation{Istituto Sistemi Complessi (ISC-CNR), Via dei Taurini 19, 00185, Rome, Italy}
\affiliation{Dipartimento di Fisica, Sapienza Universit\`a di Roma, P.le Aldo Moro 2, 00185, Rome, Italy}
    \affiliation{INFN, Unit\`a di Roma 1, 00185 Rome, Italy}
    
\author{Javier Crist\'i­n}
\affiliation{Istituto Sistemi Complessi (ISC-CNR), Via dei Taurini 19, 00185, Rome, Italy}
\affiliation{Dipartimento di Fisica, Sapienza Universit\`a di Roma, P.le Aldo Moro 2, 00185, Rome, Italy}
\email{javier.cristin@uab.cat}
 
 \author{Irene Giardina}
\affiliation{Dipartimento di Fisica, Sapienza Universit\`a di Roma, P.le Aldo Moro 2, 00185, Rome, Italy}
\affiliation{Istituto Sistemi Complessi (ISC-CNR), Via dei Taurini 19, 00185, Rome, Italy}
    \affiliation{INFN, Unit\`a di Roma 1, 00185 Rome, Italy}
    
\author{Mario Veca}
\affiliation{Dipartimento di Fisica, Sapienza Universit\`a di Roma, P.le Aldo Moro 2, 00185, Rome, Italy}


\begin{abstract}
Numerical analysis of conserved field dynamics has been generally performed with pseudo spectral methods. Finite differences integration, the common procedure for non-conserved field dynamics, indeed struggles to implement a conservative noise in the discrete spatial domain. In this work, we present a novel method to generate  a conservative noise in the finite differences framework, which works for any discrete topology and boundary conditions.  We apply it to numerically solve the conserved Kardar-Parisi-Zhang (cKPZ) equation, widely used to describe surface roughening when the number of particles is conserved. Our numerical simulations recover the correct scaling exponents $\alpha$, $\beta$, and $z$ in $d=1$ and in $d=2$. To illustrate the potentiality of the method, we further consider the cKPZ equation on different kinds of non-standard lattices and on the random Euclidean graph. 
This is the first numerical study of conserved field dynamics on an irregular topology, paving the way to a broad spectrum of possible applications.
 \end{abstract}
\maketitle

\section{Introduction} \label{Intro}

In statistical physics, field theories provide a powerful description of physical systems with many interacting degrees of freedom \cite{parisi1988statistical,kardar2007statistical,chaikin1995principles}. The evolution of the system is described via a stochastic partial differential equation for a mesoscopic field (or set of fields) $\psi$, representing the relevant quantity for the long time dynamical behavior.
Among all field theories, there is a class where conservation laws play a fundamental role and must be taken into account in the dynamical description \cite{HH1977}.  The most general form for the dynamical equation of a conserved (scalar) field $\psi(\vx,t)$ can be written as
\begin{equation}
    \frac{\partial \psi(\vx,t)}{\partial t}=-\bnabla\cdot \boldsymbol{J}  + \xi (\vx,t).
    \label{eq_field}
\end{equation}
The first term in the r.h.s of Eq.~(\ref{eq_field}) corresponds to the divergence of a current $\boldsymbol{J}(\vx,t)$, as in standard continuity equations. The functional form of $\boldsymbol J$ defines a particular dynamics, and it depends on the symmetries of the system, the nature of the interactions and the presence of constraints.
The second term in the r.h.s corresponds to a conservative noise term, whose correlator is, 
\begin{equation}
\langle \xi(\vx,t)\xi(\vx',t') \rangle =-2D \nabla^2\delta(\vx-\vx')\delta(t-t').
\label{eq_noise}
\end{equation}
 The structure of Eq. (\ref{eq_field}), together with Eq. (\ref{eq_noise}), ensures that the field $\psi$ is globally conserved. Examples of stochastic field dynamics that are described by Eqs.(\ref{eq_field}) and (\ref{eq_noise}) are the Cahn-Hilliard-Cook equation (model B) \cite{cook1970brownian,miranville2017cahn,chaikin1995principles},  Active model B \cite{stenhammar2013continuum,wittkowski2014scalar} and the conserved Kardar-Parisi-Zhang equation \cite{sun1989dynamics,krug1997origins}.
The properties of these equations have been extensively studied analytically with Renormalization Group (RG) techniques \cite{HH1977,tauber2012renormalization,lopez1999scaling,caballero2018bulk,janssen1997critical}. 
On the other hand, there has also been a great effort to investigate them numerically \cite{miranville2017cahn,wittkowski2014scalar,tiribocchi2015active}. When dealing with non-conserved fields, the main tool to perform such numerical analyses are Finite Differences (FD) methods \cite{furihata2001stable,choo1998conservative,tjhung2018cluster,emmerich2012phase,lai1991kinetic,quastel2015one}. FD methods are based on the discretization of the continuum space into a lattice, where the continuum derivatives are implemented by finite increments \cite{strikwerda2004finite,thomas2013numerical}.
FD methods, however, are problematic in the context of conserved field dynamics. More specifically, there is no obvious way to generate a conservative noise with finite differences.  This is why the standard numerical procedure for conserved field dynamics are the so-called Pseudo-Spectral (PS) methods \cite{zhu1999coarsening,basu2009scaling,wittkowski2014scalar,caballero2018strong}. PS methods alternate between integration in real space and in Fourier space \cite{fornberg1998practical,giada2002pseudospectral}, depending on the space in which each term of the equation is diagonal. The conserved noise in Eq. \eqref{eq_noise} is then easily expressed in Fourier space, where it becomes uncorrelated with variance proportional to the squared wave-vector $k^2$.

PS methods are not free of limitations. The change of basis in the PS procedure requires the knowledge of the eigenfunctions of the Laplacian operator appearing in the dynamical equation. This can be easily done when the equation is discretized on a regular lattice with Periodic Boundary Conditions (PBC) \cite{saha2020scalar}. On the contrary, it becomes highly non-trivial or even unfeasible when these conditions are not satisfied, precluding the analysis of potentially relevant applications of the considered dynamics.  Examples include the effect of arbitrary boundary conditions, the dynamical behavior on curved geometries, the presence of defects or heterogeneities. Even more generally, one might want to consider discretizations of  Eq.\eqref{eq_field}  on specific non standard topologies, which are appropriate when the underlying microscopic dynamics  occurs in non-homogeneous irregular environments. Surface growth on fractal substrates is one interesting case.

To address all such cases,  we propose in this paper a novel general scheme  to implement a conservative noise satisfying Eq.~\eqref{eq_noise} in discrete real space, which works for arbitrary topologies. To do so, we adapt the procedure presented in \cite{cavagna2023discrete}, which was developed to describe conservative fluctuations in microscopic spin dynamics. This technique allows to unambiguously study numerically equations of the class of \eqref{eq_field} with FD, and it therefore relieves from the limitations of the PS methods. To test the new scheme and to show its potentiality, we then apply it to a well-known case-study: the conserved Kardar-Parisi-Zhang (cKPZ) surface growth equation. We perform a full FD study of the cKPZ equation both for regular and non-regular lattices, showing that the method correctly reproduces the growth scaling exponents, while at the same time being completely flexible and adaptable to generic graph topologies.

The paper is organized as follows. In Section \ref{sec_noise}, we describe how to generate the conservative noise in real space.
In Section \ref{sec_ckpz}, we perform the numerical study of the cKPZ equation. We show that our numerical results are in total agreement with the theoretical predictions for the cKPZ equation in $d=1$ and $d=2$, recovering the correct scaling exponents. We also study the cKPZ equation on more complex lattices, one of them being the Euclidean random graph, a paradigmatic example of a non-regular lattice. Finally, in Section \ref{sec_conc}, we summarize our work and discuss its future applications.

\section{Conservative noise in discrete real space}
\label{sec_noise}

To discretize Eq.~\eqref{eq_field}, it is essential to address the discretization of the conserved noise with correlator \eqref{eq_noise}. The standard discrete counterpart of the continuum Laplacian  $\nabla^2$ is the discrete Laplacian operator $\Lambda_{ij}$, whose definition is given by
\begin{equation}
\Lambda_{ij} = -n_{ij}+\delta_{ij}\sum_k n_{ik} \ , 
\label{eq_lap}
\end{equation}
where $n_{ij}$ is the adjacency matrix defining the lattice's topological structure: if two sites are connected with each other $n_{ij}=1$, otherwise $n_{ij}=0$. Notice that the discrete Laplacian is a positive-definite matrix, i.e. $\Lambda \sim -a^2 \nabla^2$ ($a$ being the lattice spacing, that we set equal to $1$ in the following).
The discrete version of Eq.~(\ref{eq_noise}) then reads
\begin{equation}
   \langle \xi_i(t) \xi_j(t') \rangle = 2D \Lambda_{ij} \delta(t-t').
   \label{eq_no_disc}
\end{equation}
The question is at this point of how to generate a discrete noise term $\xi_i(t)$ satisfying such a non-trivial correlator.
In analogy with the continuum case, the natural solution would seem to build $\xi_i(t)$ as  the discretized divergence of some white noise, in such a way as to get back the Laplacian when we take the correlator, i.e. 
\begin{equation}
\xi_i(t) = [\bnabla \cdot {\boldsymbol{\eta}}]_i \ ,
\end{equation}
 (where the r.h.s. is intended in discretized form). This intuition is correct, but it has to be carefully implemented. The gradient operator in continuum space in fact depends on space in two different ways: the point where it is evaluated, and the directions along which variations are computed (corresponding to the different components of the gradient vector). When considering a discrete version of it, given a site $i$, one needs to compute finite differences with nearby sites to reproduce the possible directions of the continuum case. How to do this in a consistent way strongly depends on the structure of the discretized lattice. For a regular lattice in dimension one, there is only one possible direction, and one can for example assume $\xi_i(t) =  \eta_{i+1}(t)- \eta_i(t)$. 
 If the $\{\eta_i\}$ are random white variables with $\langle \eta_{i}(t) \eta_{j}(t')\rangle= 2D \delta_{ij}\delta(t-t')$, and PBC are considered, it is easy to show that Eq. \eqref{eq_no_disc} is then satisfied. Alternatively, one can use a symmetrized combination of white noises $\xi_i(t) =  (1/2) [\eta_{i+1}(t)- \eta_{i-1}(t)]$, as done in  \cite{nardini2017entropy}, and recover Eq. \eqref{eq_no_disc} but with a different definition of discrete Laplacian\footnote{ In  \cite{nardini2017entropy} the authors define a discretized Laplacian that involves only second nearest-neighbors rather than just the first nearest-neighbors, as in \eqref{eq_lap}. 
  }.
   The same kind of argument can also be adapted to square lattices in larger dimensions. However, when considering more complex, even regular, lattices 
it becomes quite tricky to define the appropriate combinations of finite differences between sites (i.e. a proper definition of the discrete gradient), such that Eq. \eqref{eq_no_disc} holds.
How to generalize to irregular or random lattices is far from clear.

This problem is in fact well-known in the context of graph theory, where it has been solved in an elegant way by a change of perspective. The crucial observation is that in a generic graph the relevant notion of distance is defined in terms of the links between nodes. Given a site (i.e. a node in the graph), the nearest neighbors are the nodes directly connected to it, and 
the minimal variations on the graph occur along the links  defining such connections. This suggests that a convenient definition of gradient would be in the space of sites (specifying where the gradient should be computed) and links (specifying the `directions' along which variations should be considered).  This idea is captured by the concept of {\it incidence matrix} \cite{gross2018graph}, which indeed represents the standard implementation of derivatives in graph theory.

 \subsection{The incidence matrix}\label{sec:incidence}
\begin{figure}[t]
\centering
\includegraphics[width=0.45\textwidth]{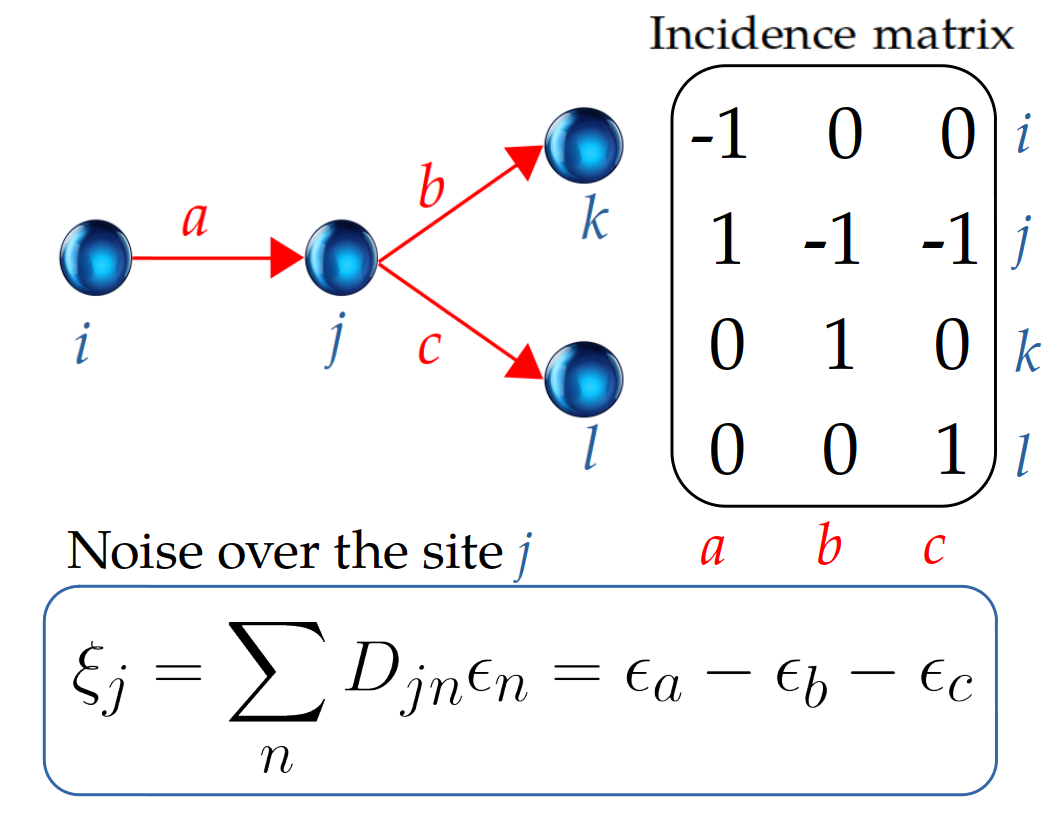}
\caption{Schematic depiction of how the conservative noise is generated for a non-regular lattice consisting of four sites. }
\label{fig:sch}
\end{figure}

 The incidence matrix $D$ is defined in the space of sites and links of the lattice, rather than of sites only. Let us label the sites of the lattice with $\lbrace i,j,\dots\rbrace$ and the links with $\lbrace a, b,\dots\rbrace$. $D_{ia}$ is constructed as follows: after arbitrarily assigning a direction to each link $a$, we set $D_{ia}=+1$ if $i$ is at the end of $a$,  $D_{ia}=-1$ if $i$ is at the origin of $a$, and $D_{ia}=0$ if site $i$ does not belong to $a$. We provide in Fig. \ref{fig:sch} an example of how the incidence matrix $D$ looks like for a non-regular lattice consisting of four sites. %

The `derivative' of a generic funtion $\psi = \{\psi_i\}$  along link  {\it a}   (i.e.  the discrete gradient) is then defined as
\begin{equation}
\left [ \nabla \psi \right ]_a = \sum_{{\rm sites}\ i} D_{ia} \psi_i \ ,
\label{discgrad}
\end{equation}
 This notion of derivative correctly reproduces the main features of the continuous one. For example, since by construction $\sum_i D_{ia}=0$, the derivative of a constant function is zero, as it should. The arbitrariness of the assignment of directions to the links reflects  the inevitable arbitrariness of defining a derivative on a regular discrete lattice (e.g. as a forward or backward finite difference). However, the definition in the link space automatically ensures some important properties for arbitrary graph topologies. In particular,  a crucial property of the incidence matrix is that its square over the links is equal to the discrete Laplacian, 
\begin{equation}
\sum_{\mathrm{links\ } a} D_{ia}D^\mathrm{T}_{aj} = \Lambda_{ij}  \ .
\label{zimbra}
\end{equation}

Finally, and perhaps more importantly for our purposes, the incidence matrix also provides a natural definition of divergence in one point. Given a generic function defined on the links $\boldsymbol{\phi}=\{\phi_a\}$, the divergence of such function at point $i$ is given by 
\begin{equation}
[\bnabla \cdot \boldsymbol{\phi}]_i= - \sum_{{\rm links} \ a} D_{ia} \phi_a \ .
\label{div}
\end{equation}

The minus sign in the definition ensures that the usual convention for the sign of the divergence operator is recovered in the continuum limit.
Besides, in this way we have $\bnabla\cdot \bnabla \psi=\nabla^2 \psi\to - \sum_a D_{ia} \sum_j D_{ja} \psi_j=-\Lambda_{ij}\psi_j$, consistently with the definition and sign of the Laplacian matrix.

\subsection{From noise on the sites to noise on the links}\label{sec:noiselink}
The incidence matrix formalism  and Eqs.~\eqref{zimbra} \eqref{div} immediately suggest how to build  a conservative noise in a discrete spatial domain. Picking up the idea of writing $\xi_i$ as the divergence of a standard white noise  it is now clear that we must switch from a white noise defined on the sites to a white noise defined on the links.  More precisely, let us define on each link $a$  a standard $\delta$-correlated Gaussian noise, $\epsilon_a$, with variance, 
\begin{equation}
\langle {\epsilon}_a(t) {\epsilon}_b(t') \rangle = 2D \;  \delta_{ab}  \, \delta(t-t') \ .
\label{pino}
\end{equation}
The site conserved noise can finally be constructed as,
\begin{equation}
\xi_i(t)=   \sum_a D_{ia} \, \epsilon_a(t) =- [\bnabla \cdot \boldsymbol{\epsilon} ]_i  \ ,
\label{zumba}
\end{equation}
where we take minus the divergence in order to have a positive sign in the discrete expression. The noise defined in Eq.~\eqref{zumba} has an immediate interpretation,  i.e. it is the sum of all the link noises incident on site $i$, and it can be easily and unambiguously generated for any kind of discrete lattice.

The variance of this new noise can be immediately computed,
\begin{eqnarray}
 \langle{\xi}_i(t) {\xi}_j(t') \rangle &=&
\sum_{ab} D_{ia} D_{jb} \, \langle \epsilon_a(t) \epsilon_b(t') \rangle
\nonumber
\\
 &=& \sum_{a} D_{ia} D^\mathrm{T}_{aj} \, 2D \, \delta(t-t') 
\nonumber
 \\
&=&2D \;   \Lambda_{ij}  \, \delta(t-t') \ .
\end{eqnarray}
We therefore recover the desired correlator \eqref{eq_no_disc}. 
Because by construction  $\sum_i D_{ia}=0$, we have that the sum over all the sites is
\begin{equation} 
\sum_i \xi_i = 0 \ ,
\end{equation}
which explicitly shows that the noise is globally conserved. 

\subsection{Multiplicative conservative noise}\label{sec:mnoise}
The formalism developed in the previous sections can be generalized to also address conserved stochastic field equations with multiplicative noise. In this case, the conserved noise appearing in Eq.\eqref{eq_field} depends on the field itself, i.e.

\begin{equation}
 \xi (\vx,t) = -\bnabla\cdot \left ( f\left [ \psi(\vx,t)\right ]  \boldsymbol{\epsilon} (\vx,t) \right ),
 \label{noisem}
\end{equation}
where $ \boldsymbol{\epsilon} (\vx,t)$ is a  Gaussian white noise with variance $\langle \boldsymbol{\epsilon} (\vx,t) \cdot \boldsymbol{\epsilon} (\vx^\prime,t)\rangle = 2 d D \boldsymbol{\delta}(\vx-\vx^\prime)\delta(t-t^\prime)$, $d$ is the space dimension, and $ f[\psi]$ is a scalar function of the field ($f=1$ reproducing the additive noise case discussed so far)
\footnote{
 In Eq.~\eqref{noisem} we use minus the divergence, in analogy with Eq.\eqref{zumba}. In both equations, being $\boldsymbol{\epsilon}$ a random white noise, the sign is completely irrelevant and it can be chosen for convenience.
 }.
 Relevant examples can be found in the Dean-Kawasaki equation and generalizations \cite{dean1996langevin,kawasaki1993relaxation,Lefevre_2007}, with potential interesting applications to reaction-diffusion processes, stochastic density functional theory and macroscopic fluctuation theory  \cite{Lefevre_2007,Demery_2016,MFT,krapivsky2015tagged}. To discretize Eq.~\eqref{noisem}, we first notice that \begin{equation}
\bnabla\cdot \left ( f   \boldsymbol{\epsilon} \right ) = f \bnabla\cdot \boldsymbol{\epsilon} + \bnabla f \cdot \boldsymbol{\epsilon} \ .
\label{rule}
\end{equation}
To implement this expression on a generic discrete lattice we proceed as before and introduce a white noise defined on the links, i.e. $\boldsymbol{\epsilon}(\vx,t)\to \left \{ \epsilon_a \right \}$. The first term in the r.h.s. of Eq.~\eqref{rule} can then be immediately discretized using the graph divergence defined in \eqref{div}. The second term is more tricky, as it requires a proper definition of local scalar product between two link dependent functions, i.e. $[\nabla f]_a=\sum_i D_{ia}f_i$ and $\epsilon_a$. How to do that in a consistent way is explained in Appendix \ref{app:B}. The result is
\begin{equation}
\xi_i =   f_i \sum_a D_{ia} \epsilon_a -\frac{1}{2} \sum_{a\ni i} \sum_j D_{ja} f_j \ \epsilon_a \ ,
\label{discm}
\end{equation}
where $f_i=f[\psi_i]$
and the second sum is restricted to the links $a$ incident on site $i$. 
It can be easily verified that this discrete noise is conserved. Indeed, summing over sites both members of Eq.~\eqref{discm}, and considering that $\sum_i \sum_{a\ni i} = 2 \sum_a $, we immediately get $\sum_i \xi_i =0$.

\section{The conserved KPZ equation}
\label{sec_ckpz}

One interesting model that belongs to the class described by Eqs.~\eqref{eq_field} and \eqref{eq_noise} is the conserved Kardar-Parisi-Zhang equation \cite{sun1989dynamics} (cKPZ), a conservative variant of the more widely known Kardar-Parisi-Zhang equation \cite{kardar1986dynamic,corwin2012kardar}. It describes the dynamics of a growing surface under the constraint that the total surface height is conserved. The cKPZ equation reads 
\begin{equation}
\frac{\partial h(\vx,t)}{\partial t} =  - \nabla^2(\nu \nabla^2 h(\vx,t) + \lambda (\nabla h)^2) +\xi (\vx,t),
     \label{ckpz}
\end{equation}
where $h(\vx,t)$ represents the surface height field. The main properties of the surface are described by the height fluctuations and, in particular, by the average surface width $W$,
\begin{equation}
    W(L,t)=\Big\langle \frac{1}{L^d} \int d\vx \ \lbrace h(\vx,t)-\Bar{h}(t) \rbrace^2 \Big\rangle^{1/2},
\end{equation}
where $\Bar{h}(t)=(1/L^d) \int d\vx h(\vx,t)$ is the average sample height, $d$ is the dimension of the surface space and $L$ is the size of the system. According to dynamic scaling theory, $W$ follows the Family-Vicsek scaling relation \cite{family1985scaling}
\begin{equation}
    W(L,t) \sim L^\alpha f(t/L^{z}) \ ,
    \label{scaling}
\end{equation}
where the scaling function $f(x)$ approaches a constant for $x \gg 1$, while $f (x) \sim x^\beta$ for $x \ll 1$ with $z = \alpha / \beta$. The exponents $\alpha$,
$\beta$, and $z$ are called the roughness, the growth, and the dynamic exponent, respectively. The cKPZ equation describes an inherently out-of-equilibrium dynamics, as it cannot be derived from a Hamiltonian. A Renormalization Group (RG) analysis  \cite{sun1989dynamics,janssen1997critical,mukherjee2021conserved} has shown that the exponents $\alpha$, $\beta$ and $z$ in $d$ dimensions are:
\begin{equation}
\alpha= \frac{\epsilon}{3} \: \:\:  ; \: \:\: z=4-\frac{\epsilon}{3} \: \:\:  ; \: \:\: \beta=\frac{\alpha}{z}=\frac{\epsilon}{12-\epsilon},
\label{ckpz_sca}
\end{equation}
where $\epsilon=d_c-d$, with upper critical dimension $d_c=2$ \cite{janssen1997critical,mukherjee2021conserved}. We remark that the conservative noise has a fundamental role in determining these exponents. Indeed, when a non-conservative noise is considered the universality class changes to the Lai-Das-Sarma one \cite{lai1991kinetic}.

Numerical integration of the cKPZ equation has been performed in several works, and correctly reproduces the predicted scaling exponents \cite{caballero2018strong,chakrabarti1990computer}. However, past numerical studies have only been performed using PS methods,  for the reasons discussed in the Introduction. This restricts them to the case of regular lattices with PBC. 

For more general cases, the method we propose - where the conservative noise is implemented within a FD framework - provides a natural way to numerically solve the cKPZ equation. Moreover, as an additional advantage, FD  numerical integration of a lattice with $N$ sites requires $\mathcal{O}(N)$ operations for each timestep, while the PS method requires $\mathcal{O}(N\log{}N)$ operations \cite{gallego2007pseudospectral}. In the next sections, we discuss how to efficiently integrate the equation via FD, and we study it on several  kinds of lattices.
 

\subsection{The discretized cKPZ equation}\label{sec:grad2}
Let us now proceed with the discretization of the cKPZ equation. In the previous section we already discussed how to treat the conserved noise. The linear term in the r.h.s. of Eq.~\eqref{ckpz} is straightforward to deal with, as we can use the definition of the Discrete Laplacian given in Eq.~\eqref{eq_lap}. The remaining non-linear  term involves a squared gradient $(\nabla h)^2$. We therefore encounter, again, the problem of choosing an appropriate representation of the gradient operator in discrete space.  For square discrete lattices, different definitions have been used so far in the literature \cite{dasgupta1997,beccaria1994numerical,buceta2005generalized}, where different prescriptions are considered for taking finite differences between neighboring sites (e.g. using a forward or backward or symmetric rule). As long as the considered system is homogeneous, small local differences in the definition of the discrete gradient should not change the large scale behavior. However, as also discussed before, we seek a discrete representation of all the operators appearing in the equation that can be generalized to more complex (even strongly heterogeneous) lattices and boundary conditions. Besides, it would be desirable that all the terms in the equation are treated in a consistent way, without relying on any arbitrary choice.

To address this issue, we propose a general and simple method to write the non-linear term in a way that is automatically consistent with the Laplacian, and that can be unambiguously discretized on any lattice.

To do this, we note that for any given scalar field $f$, the squared gradient can be expressed as
\begin{equation}
    (\nabla f)^2= \bnabla \cdot (f \bnabla f)-f \nabla^2f \ .
\end{equation}
The term inside the divergence can be written as
\begin{equation}
    f \bnabla f= \frac{1}{2}\bnabla (f^2) \ ,
\end{equation}
and we then get for the squared gradient
\begin{equation}
    (\nabla f)^2= \frac{1}{2} \nabla^2 (f^2)-f\nabla^2 f.
    \label{formula-mario}
\end{equation}
From Eq.~\eqref{formula-mario} we see that the squared gradient can be rewritten in terms of the Laplacian operator only, its 
discretization is therefore uniquely defined by the corresponding discrete Laplacian $\Lambda$, i.e.
\begin{equation}
\left [(\nabla f)^2 \right ]_i = -  \frac{1}{2}  \sum_{k}  \Lambda_{ik} f_k^2  +  f_i \sum_{k}  \Lambda_{ik} f_k \ .
\label{grad2disc}
\end{equation} \
This expression can be used for any topology without ambiguities, which is a crucial feature if one is interested in discretizing the continuous equation on non trivial lattices. 

We note that  this fairly simple argument leads to the same discretization of the non-linear term as the one obtained in  \cite{wio2010discretization} in the context of the standard KPZ equation. In that work, the authors exploited the well-known mapping of KPZ to a diffusion equation with multiplicative noise \cite{bertini1997stochastic} to show that the discrete implementation of the squared gradient is actually constrained by the definition of the discrete Laplacian. The mapping to a diffusion equation does not hold in the conserved case, but relations \eqref{formula-mario} \eqref{grad2disc} are general and they can be applied to all cases where squared gradients appear in the dynamical equations\footnote{Using similar kinds of mathematical relations, it is possible solve the ambiguity issue for the discretization of more complicated non-linear terms appearing in other equations, like in \cite{caballero2018strong}.}.

We therefore write the general discrete cKPZ equation for any lattice as
\begin{equation}
    \frac{d h_i}{dt}=\sum_{k,j} \Lambda_{ik}\Lambda_{kj}[-\nu h_j-\frac{\lambda}{2}h_j^2+\lambda h_kh_j] + \xi_i,
    \label{eq_disc}
\end{equation}
with noise correlator given by Eq.~\eqref{eq_no_disc}, that we rewrite here for convenience
\begin{equation}
   \langle \xi_i(t) \xi_j(t') \rangle = 2D \Lambda_{ij} \delta(t-t').
   \label{eq_no_disc1}
\end{equation}

To conclude this section, we note that the same discrete equation for an arbitrary graph can be obtained - even though with considerable  more algebra - using the definition of incidence matrix given in the previous section, and writing the squared gradient in terms of discrete derivatives along the links (see App.\ref{app:B}). This provides a further consistency check on the whole discretization procedure.

\begin{figure*}[t]
\centering
\includegraphics[width=0.9\textwidth]{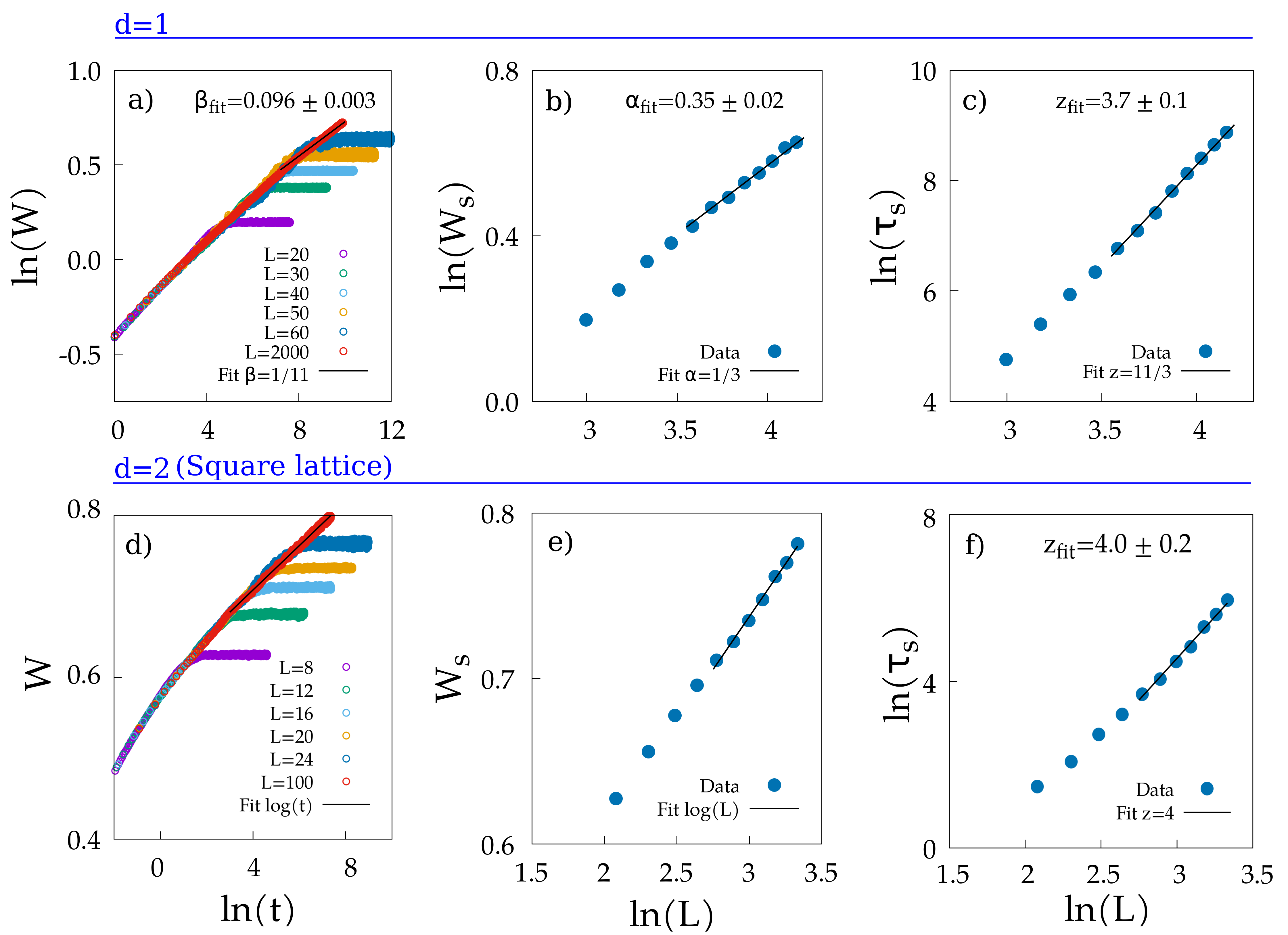}
\caption{Numerical results in $d=1$ (upper row) and $d=2$ for a square lattice (lower row). a) Interface width $W$  as a function of time for different system sizes $L$.  The black line corresponds to the best fit of the $L=2000$ curve with $\beta=\frac{1}{11}$. The best free fit gives an exponent $\beta_{fit} =0.096 \pm 0.003$. b) Saturation width $W_s$ for different system sizes $L$. The black line corresponds to best fit with $\alpha=\frac{1}{3}$. The best free fit gives an exponent $\alpha_{fit} = 0.35 \pm 0.02$. c) Saturation time $\tau_s$, estimated from the curves in (a) (see text), for different system sizes $L$. The black line corresponds to the best fit with $z=\frac{11}{3}$. The best free fit gives an exponent $z=3.7 \pm 0.1$. d) Surface width $W$ as a function of time  for different system sizes $L$. The black line corresponds to the logarithmic fit. e) Saturation width $W_s$ for different system sizes $L$. The black line corresponds to the logarithmic fit. f) Saturation time $\tau_s$, for different system sizes $L$. The black line corresponds to the best fit with $z=4$. The best free fit gives an exponent $z=4.0 \pm 0.2$. The number of simulated samples goes from $10000$ for $L=20$ to $500$ for $L=64$, and $20$ for $L=2000$ in $d=1$; and  from $10000$ for $L=8$ to $500$ for $L=28$, and  $5$ for $L=100$ in $d=2$.}
\label{fig:1}
\end{figure*}

\subsection{Validation on Cartesian lattices}
\label{sec:validation}
Given Eqs.~\eqref{eq_disc}, \eqref{eq_no_disc1}, and the way to generate the conservative noise described in Sec.~\ref{sec:noiselink}, we can perform numerical simulations of the cKPZ equation working exclusively in real space, and on any kind of underlying lattice. As a starting point, though, we wish to check that the procedure outlined so far correctly works on known cases. We therefore initially consider the same kind of standard topologies where previous numerical analysis have been performed. Results are displayed in Fig.~\ref{fig:1} and show that our method provides estimates of the scaling exponents that are in full agreement with the RG predictions  \eqref{ckpz_sca} and with the numerical results obtained with PS methods in $d=1$ and $d=2$ \cite{chakrabarti1990computer,caballero2018strong}.
 
We considered a regular lattice in $d=1$, and a regular square lattice in $d=2$. In both cases, the lattice spacing has been set to $a=1$. The size of the system is defined by the length $L$. The number of sites is therefore, respectively, $N=L$ in $d=1$ and $N=L^2$ in $d=2$.  To perform the numerical integration in time of Eq.\eqref{eq_disc} we used an Euler integration scheme with time step $\Delta t = 2 \times 10^{-3}$ and PBC. The initial spatial distribution of the field $h_i$ has been taken randomly yet ensuring that its total sum is $0$. The parameters have been chosen as $\nu=0.5$ and $\lambda=1$, in analogy with previous numerical works \cite{caballero2018strong,chakrabarti1990computer}.

\begin{figure*}[t]
\centering
\includegraphics[width=1\textwidth]{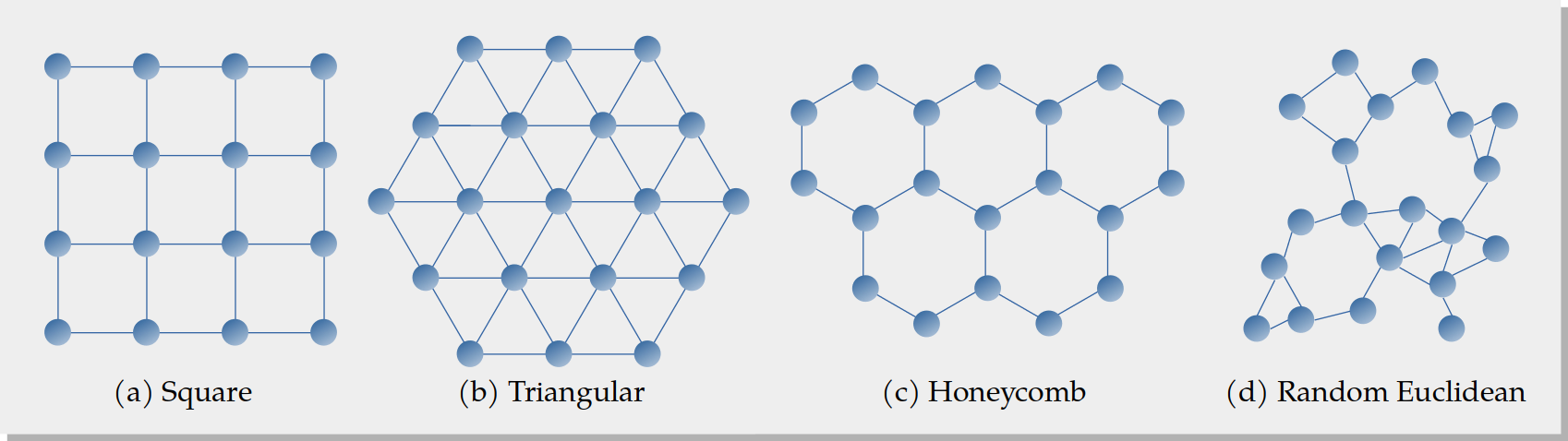}
\caption{Visual representation of the $4$ lattices in $d=2$ that have been used to numerically study the cKPZ equation.
}
\label{fig:latt}
\end{figure*}

As discussed above, the quantity of interest is the average surface width $W(L,t)$, whose behavior is characterized by the three relevant exponents $\alpha$, $\beta$ and $z$ (see Eq.\eqref{scaling}). $W$ is displayed in Fig.~\ref{fig:1}a,d, where it is plotted as a function of time, for different system sizes. The scaling relation Eq.\eqref{scaling} implies that $W$ should initially grow with time as $t^\beta$, and then saturate over a time $\tau_s \sim L^z$ to a size dependent asymptotic value $W_s \sim L^\alpha$. This is well reproduced by our curves, which also allow to extract the three exponents. In particular, a fit of $W$ vs $t$ in the growing regime gives an estimate of $\beta$, while fits of the saturation width $W_s$ and the saturation time $\tau_s$ vs. $L$ give estimates of $\alpha$ and $z$.

For $d=1$, we performed simulations with system sizes ranging from $L=20$ to $L=64$.  
The duration of the runs (number of time-steps) has been chosen to ensure that the saturation width $W_s$ was reached. We also simulated a particularly large system,  with $L=2000$, where saturation is not reached and $W$ remains in the growing regime for the whole simulation time. This allowed to accurately estimate the exponent $\beta$, as displayed in Fig.~\ref{fig:1}a. 

For the smaller sizes, we computed the saturation width $W_s$ as the average of the stationary region in Fig.~\ref{fig:1}a, and the saturation time $\tau_s$ as the intersection between a power-law fit of the initial growth regime and the saturation width. The resulting values of $W_s$  and $\tau_s$ are displayed in Fig.~\ref{fig:1}b and c, as a function of the size $L$ of the system, in log-log scale. As clearly shown in Fig.~\ref{fig:1}, the numerical data are fully consistent with the theoretical predictions, $\alpha=\frac{1}{3}$, $\beta=\frac{1}{11}$ and $z=\frac{11}{3}$, displayed as black lines in the figures. A fit of the curves in the three top panels gives the numerical estimate $\alpha_{fit}=0.35 \pm 0.02$, $\beta_{fit}=0.096 \pm 0.003$ and $z_{fit}=3.7 \pm 0.1$.

For $d=2$, we followed a similar strategy. We performed simulations for sizes ranging from $L=8$ to $L=28$, where all the $W(t)$ curves reach the saturation value, and a larger size $L=100$ for which only the growing regime is observed. Again, this largest size is used to estimate the exponent $\beta$. Since $d=2$ corresponds to the upper critical dimension of the cKPZ equation, the theoretical predicted value for the growth exponent is $\beta=0$, which implies a logarithmic behavior. It is therefore convenient in this case to plot the $W$ vs $t$ curves in linear-log scale, and directly verify the logarithmic dependence, which we successfully do in Fig.~\ref{fig:1}d. The saturation value $W_s$ is computed as the average value of $W(t)$ in the stationary time regime, as before. For $d=2$ the scaling relation implies $W_s(L)\sim \log(L)$, corresponding to a theoretical value of the roughness exponent $\alpha=0$. In Fig.~\ref{fig:1}e, we plot $W_s$ vs. $L$ in linear-log scale, and show that the numerical data perfectly satisfy the expected behavior. Finally, we compute the saturation time $\tau_s$ as the intersection between the logarithmic fit of the initial regime in Fig.~\ref{fig:1}d and the $W_s$ value. $\tau_s$ is shown in Fig.~\ref{fig:1}f as a function of $L$ in log-log scale. A fit of the data gives  $z_{fit}=4.0 \pm 0.2$, fully consistent with the theoretical prediction $z=4$.


\begin{figure*}[t]
\centering
\includegraphics[width=0.88\textwidth]{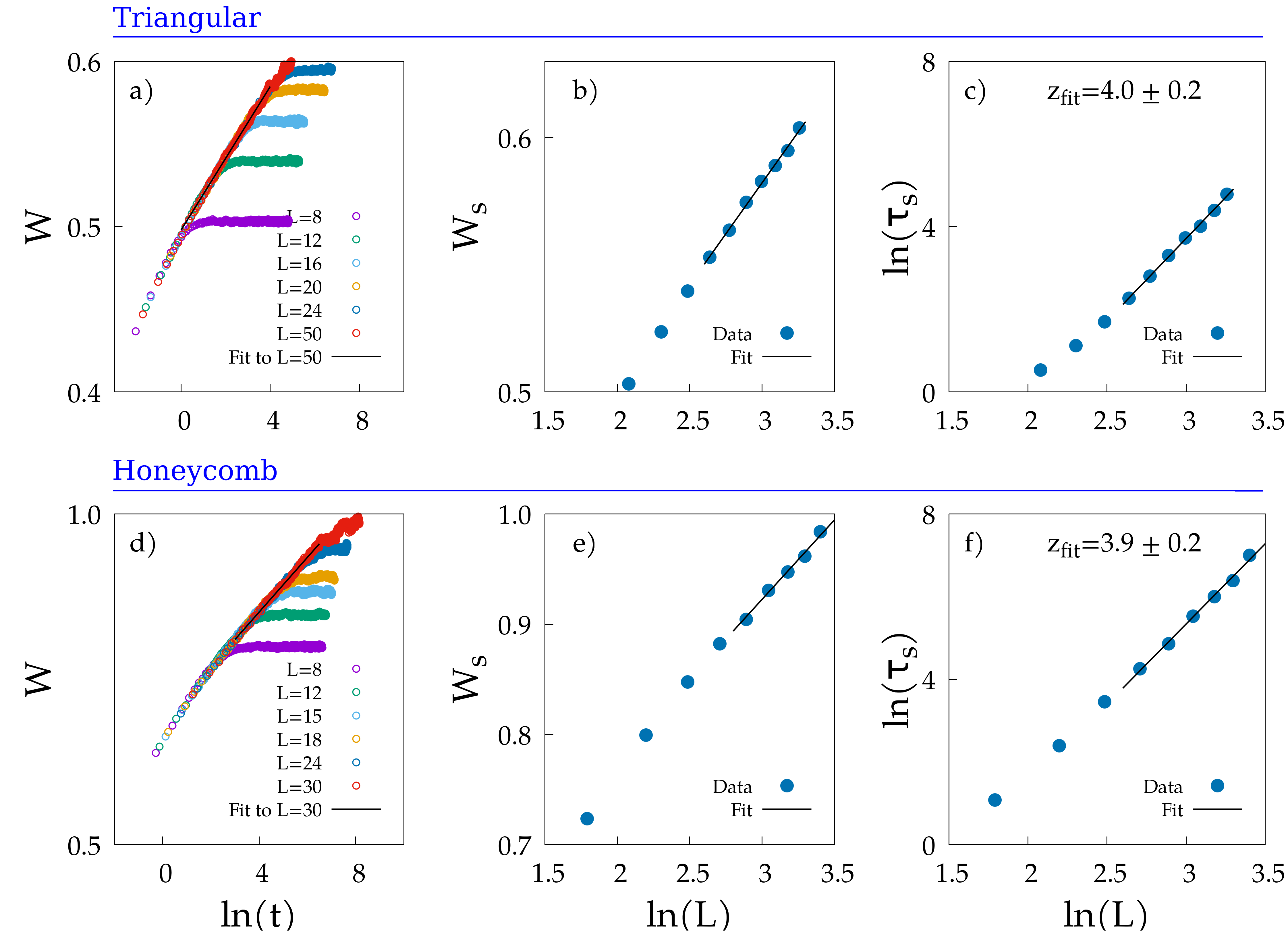}
\caption{Numerical results for the triangular lattice (upper row) and the honeycomb lattice (lower row).  a) and d) Surface width $W$ as a function of time for different system sizes $L$. The black line corresponds to the best logarithmic fit. b) and e) Saturation width $W_s$ as a function of the system size $L$. The black line corresponds to best logarithmic fit. c) and f) Saturation time $\tau_s$, estimated from the curves in a) and d), as a function of the system size $L$. The black line corresponds to the best fit with $z=4$. The best free fit gives an exponent $z=4.0 \pm 0.2$ for the triangular lattice, and  $z=3.9 \pm 0.2$ for the honeycomb one.  For both lattice types the number of samples goes from $10000$ for the smallest size, to $500$ for the largest size reaching saturation, while $5$ samples have been used for the simulations where saturation is not reached (red curves in panels a) and d)).
}
\label{fig:num_latt}
\end{figure*}

\subsection{Triangular and honeycomb lattices}
In the previous Section we have shown that our  method can be easily applied to square lattices in $d=2$, correctly recovering the results found in the past with PS \cite{caballero2018strong,chakrabarti1990computer}. 
Still, among regular lattices, the square one is the simplest case, as its base vectors are orthogonal. This is not true for other regular, yet more complex lattices.  Paradigmatic examples are the triangular and the honeycomb lattices (see  Fig.\ref{fig:latt}b,c).  Already in the triangular lattice standard directional derivatives do not have any straightforward expression, as they should involve multiple neighbouring sites. The honeycomb case is even more complicated, since the spatial inversion symmetry does not hold. 
In both these examples, standard FD methods would require {\it ad hoc} complicated prescriptions. On the contrary, our method to generate the conservative noise works exactly as in the square lattice case, precisely because it is defined on links (irrespective of their spatial structure). Besides, our trick to express the squared gradient in terms of the discrete Laplacian $\Lambda_{ij}$ provides a natural way to solve the ambiguity of the directional derivative.

We have therefore studied numerically Eqs.~\eqref{eq_disc}, \eqref{eq_no_disc1} on the triangular and honeycomb lattices. For the triangular case, we performed simulations of system sizes ranging from $L=8$ to $L=28$ all reaching saturation within simulation time, together with a very large size $L=50$, where saturation is not reached. Similarly, for the honeycomb case, we considered system sizes ranging from $L=9$ to $L=24$ all reaching saturation, and $L=30$, where saturation is not reached. We then followed the same procedure detailed for the $d=2$ square lattice in the previous section to analyze data and estimate the exponents. Results are displayed in Fig.~\ref{fig:num_latt}. We note that both lattices are bi-dimensional and regular, with local connections between sites. We therefore expect that the large scale behavior is exactly the same as for the $d=2$ square lattice. Indeed this is what we find: a logarithmic growth is observed for $W(t)$ in the initial regime (Fig.~\ref{fig:num_latt} a,d), the saturation width $W_s$ depends logarithmically on the system size (Fig.~\ref{fig:num_latt} b,e), and the dynamic exponent $z$ is very close to the theoretical prediction $z=4$ (we find $z_{fit}=4.0 \pm 0.2$ in the triangular case, and $z_{fit}=3.9 \pm 0.2$ in the honeycomb one - see Fig.~\ref{fig:num_latt} c,f).

\begin{figure*}[t]
\centering
\includegraphics[width=1\textwidth]{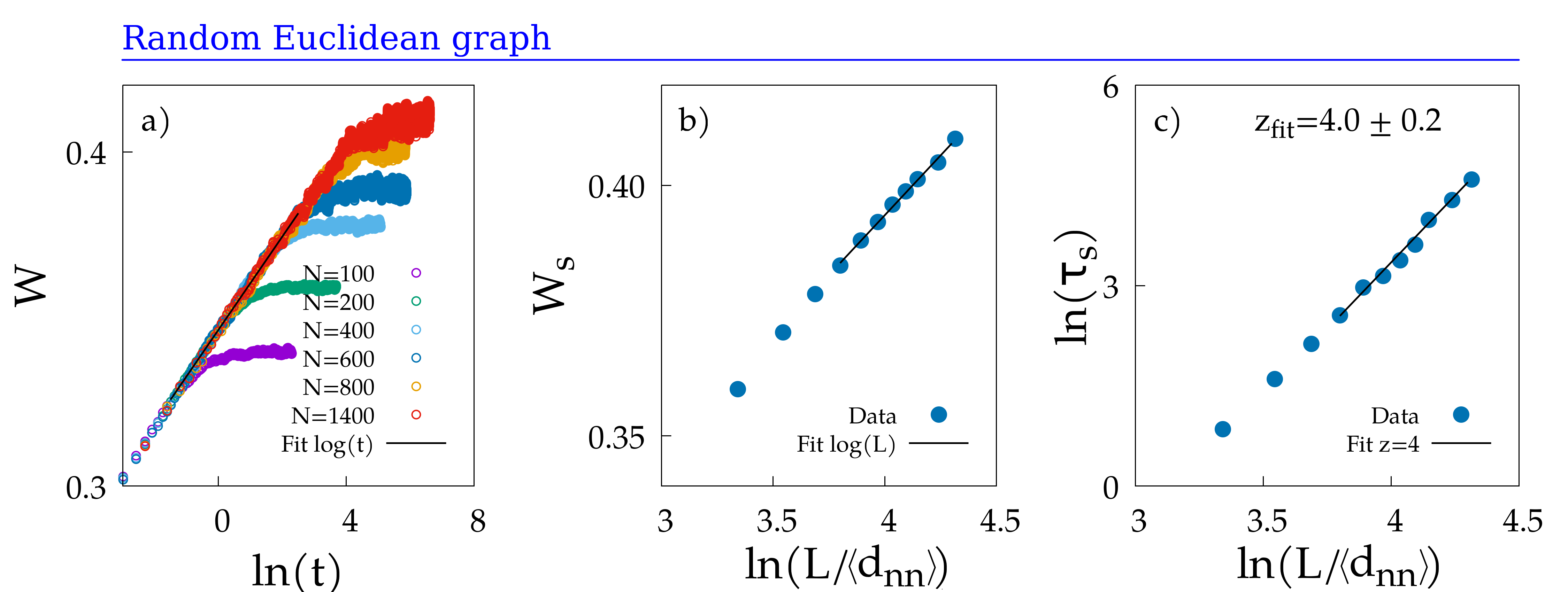}
\caption{Numerical results for the random Euclidean graph embedded in $d=2$. a) Surface width $W$ as a function of time for different system sizes 
$N$.
The black line corresponds to the best logarithmic fit of the largest size. b) Saturation width $W_s$ as a function of the normalized system size $L/\langle d_{nn}\rangle$. The black line corresponds to best logarithmic fit. c) Saturation time $\tau_s$, which has been
estimated from the curves in a), as a function of the normalized system size $L/\langle d_{nn}\rangle$. The black line corresponds to the best fit with $z=4$. The best free fit gives an exponent $z=4.0 \pm 0.2$. The number of simulated samples goes from $10000$ for the smallest size to $500$ for the largest one. The density is kept fixed to $\rho=4$, therefore for each size $L=\sqrt{N}/2$.}
\label{fig:rand}
\end{figure*}
\subsection{Random Euclidean graph}

A further step forward in the direction of more complex cases are non-regular lattices. This is the scenario where the full potentiality of our method comes into play, as no standard FD nor PS implementations are feasible. On the contrary, having defined the discretized equations for a generic graph topology, our method can address this kind of problems as easily as in the square lattice case.

To show this, we now consider a non-regular example in $d=2$, namely the Euclidean random graph \cite{dall2002random,penrose2003random} (Fig.~\ref{fig:latt}d). The definition of such graph is as follows: one randomly throws $N$ points in a square of length $L$; if two points are separated by less than a given Euclidean distance $r_c$, they are connected (having $n_{ij}=1$) otherwise they are not (i.e. $n_{ij}=0$). Contrary to a regular lattice, there is no fixed lattice spacing $a$ in this case. However, we can consider as equivalent microscopic length-scale the average nearest neighbor distance $\langle d_{\rm nn} \rangle$. More details about the Euclidean random graph are given in  App.~\ref{app:A}. Here we note that - even though this graph is not regular - the connections between sites are still local in space. 
Hence, we still expect that the large-scale phenomenology is the same as in the regular cases described in the previous sections, even although in this case the test is significantly less trivial.

 In our numerical analysis, we considered graphs with sizes ranging from $N=100$ to $N=1400$. For any given value of $N$, the length $L$ of the square containing the points is chosen to keep the density fixed, i.e. $\rho=N/L^2=4$. The connectivity distance is set to $r_c=1$. These values ensure that there is only one connected cluster and that we are away from the mean field limit, since $\pi r_c^2 \ll L^2$ for all the considered values of $N$. 
 Under these conditions, the exponents $\alpha$, $\beta$ and $z$ should be exactly equal to the ones obtained for the square regular discretization in $d=2$ (and for the triangular and honeycomb lattices). The numerical results are shown in Fig.~\ref{fig:rand} where, to compare with previous cases (where the lattice spacing is $a=1$), we use the size $L$ of the system normalized to the average nearest neighbor distance $\langle d_{\rm nn} \rangle$ (computed numerically). We recover the logarithmic growth both for $W(t)$ and $W_s$ and we get $z=4.0 \pm 0.2$ for the dynamic exponent. The fact that we obtain the same exponents in the square, in the triangular, in the honeycomb and in the random Euclidean graph is in perfect agreement with the notion of universality in the framework of the RG.

\section{Conclusions}\label{sec_conc}

We have proposed a novel method to generate a discrete conservative noise in real space, a useful tool to numerically solve conserved stochastic field dynamics with FD. The strength of our scheme lies in its simple formulation and in its generality, making it the natural way for the analysis of conserved dynamical equations on any discrete topology. 
We used the method to investigate numerically a well-known case of conserved stochastic dynamics, namely the cKPZ equation. We have shown that for the standard regular discretizations in $d=1$ and $d=2$ it recovers the correct scaling exponents predicted by RG calculations and found by PS methods. Furthermore, we extended our analysis to study the cKPZ on more complex discrete lattices, where it had never been considered before: the triangular and the honeycomb lattices, and the random Euclidean graph. This last case represents the first instance where a conserved field dynamics has been addressed on a non-regular discrete structure. Our analysis provides the expected results in terms of scaling behavior and universality of critical exponents. 

We believe that our method has a broad range of possibile applications for the analysis of conservative field dynamics; its implementation is straightforward not only for the cKPZ considered here, but also for the Cahn-Hilliard-Cook equation, for Active Model B, for the Dean-Kawasaki equation, or any other conserved equations either with additive or multiplicative noise. 
A most promising outlook is the study of conservative dynamics on complex topologies, where previous FD and PS methods cannot be applied; in these cases, link-noise is the {\it only} way to proceed. 
Some examples of physical relevance to be investigated in the future include, among others: 
conserved dynamics on a spherical surface, where any discretization involves triangular loops \cite{private}; random pinning in bulk properties, when there are isolated missing sites; 
the study of fixed boundary conditions and of the surface effects induced by them. One final case of particular interest is the study of conserved dynamics on fractal substrates.  Such structures exhibit a non-integer dimension, and it is not yet fully understood whether or not RG predictions involving an $\varepsilon$ expansion apply to this case. A numerical analysis may therefore provide useful insights for a deeper theoretical understanding of critical phenomena.

\vskip 1 cm

\section*{Acknowledgements}

This work was supported by ERC grant RG.BIO (n. 785932), and by grants PRIN-2020PFCXPE and FARE-INFO.BIO from MIUR. We thank G. Pisegna, F. Ricci-Tersenghi and M. Scandolo for fruitful discussions, and D. Venturelli for several interesting comments and for suggesting us the case with multiplicative noise.

\bibliography{main_revised}

 \appendix 
\begin{figure*}[h!]
\centering
\includegraphics[width=0.75\textwidth]{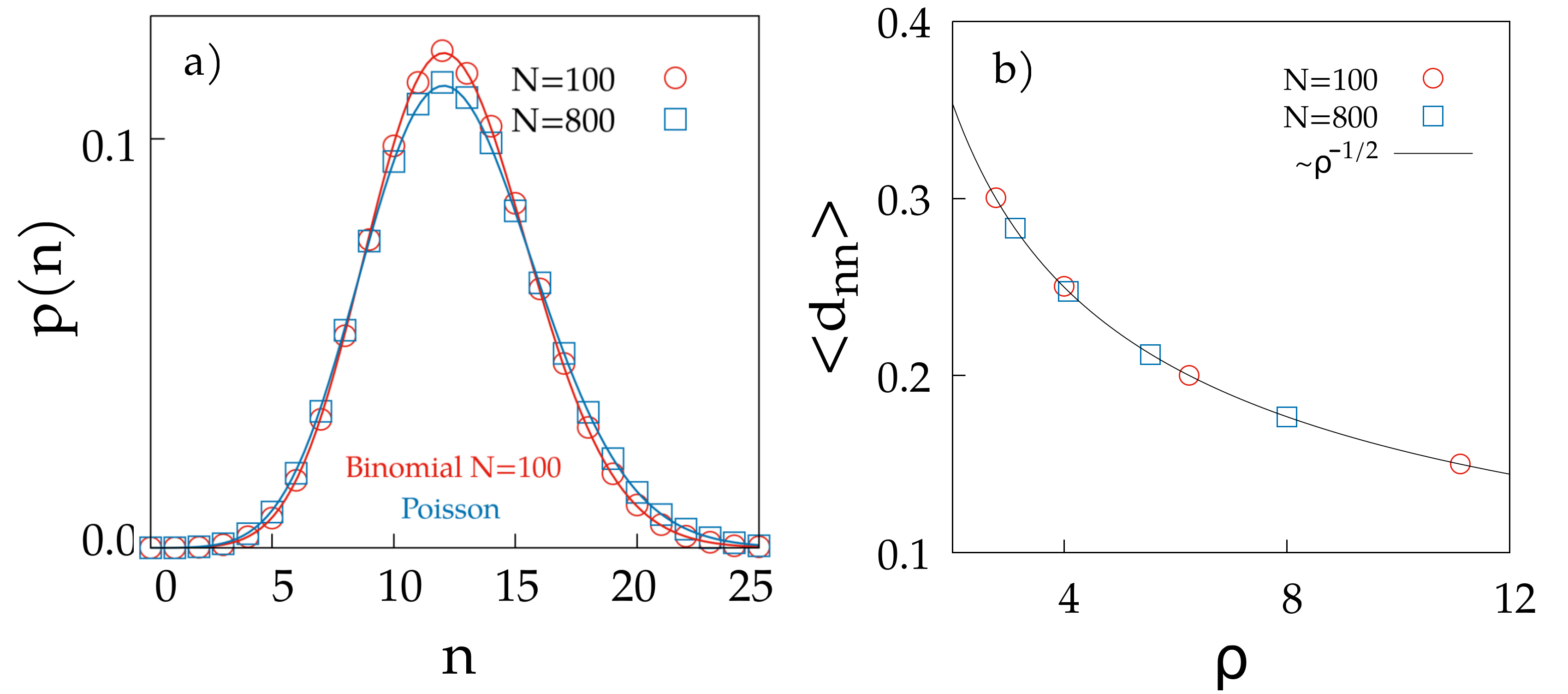}
\caption{a) Distribution of the number of interacting neighbours in the random Euclidean graph for two different sizes, with  density $\rho=\frac{N}{L^2}=4$ and $r_c=1$ (the values used in the main text). The red line corresponds to the Binomial distribution (with $N=100$ and $r=\pi r_c^2/L^2$ - see text) and the blue line corresponds to the Poisson distribution (with $\langle n \rangle = \rho \pi r_{c}^2$ - see text) . b) Average distance of the nearest neighbour as a function of the density $\rho$.}
\label{fig:rand1}
\end{figure*}

\section{Local scalar products on the graph via the incidence matrix}\label{app:B}
In stochastic field equations there are sometimes terms involving a scalar product between the gradient of the field and some other vectorial function, evaluated at the same point in space. In this paper we encountered two such instances: the term $\bnabla f \cdot \boldsymbol{\epsilon}$  appearing in stochastic equations with multiplicative noise (see Sec.~\eqref{sec:mnoise}); and the $(\nabla h)^2$ term present in the current of the cKPZ equation (see  Sec.~\eqref{sec:grad2}). In this appendix we illustrate how to discretize in a coherent way such contributions using the incidence matrix formalism.

As discussed in Sec.~\ref{sec:incidence}, on a graph the derivatives  are defined along links, rather than along  cartesian directions, e.g. $[\nabla f]_a = \sum_i D_{ia} f_i$ and $[\nabla h]_a = \sum_i D_{ia} h_i$. The noise also becomes a random function $\epsilon_a$  on the links rather than on the sites. To implement the discrete version of the terms mentioned above, we therefore need to construct a scalar product between link functions. More specifically, what we seek is a definition of scalar product that reproduces the features of the continuous one. To this end, it must i) be local; ii) be consistent with the continuum limit on a regular lattice. Following these guidelines, given two generic functions ${\bf g}=\{g_a\}$ and ${\bf l}=\{l_a\}$ on the links, we define the scalar product between them at site $i$ as
\begin{equation}
[\boldsymbol{g}\cdot \boldsymbol{l}]_i  = \frac{1}{2} \sum_{a \ni i} g_a l_a
 \label{sprod}
\end{equation}
where the sum is restricted to the links $a$ incident on site $i$ (to ensure locality), and the factor $1/2$ is due to consistency with the continuum limit from a regular discretization (where cartesian coordinates are half the number of the connected neighbors). Let us now apply this definition to the two cases encountered in this paper.

\vskip 0.5 cm
\noindent 
$\bullet$ {\it Discretization of the multiplicative noise term}

When addressing the case of multiplicative noise, a term $\bnabla f \cdot \boldsymbol{\epsilon}$ appears in the stochastic part of the dynamical equation, where $f=f[\psi]$ is a scalar function of the field and $\boldsymbol{\epsilon}$ is a white delta correlated Gaussian noise (see Eq.~\eqref{rule}). Following definition \eqref{sprod}, on a discrete lattice this product becomes
\begin{equation}
[(\bnabla f)\cdot \boldsymbol{\epsilon}]_i  = \frac{1}{2} \sum_{a \ni i} [\nabla f]_a \epsilon_a
= \frac{1}{2} \sum_{a \ni i} \sum_j D_{ja} f_j \epsilon_a \ ,
 \label{sprod1}
\end{equation}
which is precisely what appears in the r.h.s. of Eq.~\eqref{discm}.

\vskip 0.5 cm
\noindent
 $\bullet$ {\it The squared gradient term in the cKPZ equation}

In Sec.~\ref{sec:grad2} we discussed how to express the $(\nabla h)^2$ term of the cKPZ equation in terms of the Laplacian, leading to an unambiguous discretization. Here we address the same problem within the formalism of the incidence matrix. Given that $(\nabla h)^2 = \bnabla h \cdot \bnabla h$,  we can directly apply Eq.\eqref{sprod} and we get,
\begin{align}
[(\nabla h)^2]_i &= \frac{1}{2} \sum_{a \ni i} [\nabla h]_a [\nabla h]_a = \nonumber \\
&= \frac{1}{2} \sum_{a \ni i} \sum_j D_{ja} h_j \sum_k D_{ka}h_k \ .
\label{nabla2}
\end{align}
Expression \eqref{nabla2} can be rewritten in a clearer way by exploting the definition and the properties of the incidence matrix. In particular we have 
\begin{equation}
D_{ia} D_{ja} = 
\begin{cases}
\ \ 0  &\quad  {\rm if} \quad  i,j \notin a \\
-1 &\quad  {\rm if}\quad   i,j \in a \  \  \wedge \ i\ne j\\
\ \ 1 & \quad  {\rm if}\quad i,j \in a \ \ \wedge  \  i=j\\
\end{cases}
\label{dprop}
\end{equation}
Then we have
\begin{align}
&\frac{1}{2} \sum_{a \ni i} \sum_j D_{ja} h_j \sum_k D_{ka}h_k  = \nonumber \\
& =  \frac{1}{2} \sum_a D_{ia}^2 \left [ \sum_{k\ne j}  D_{ja}  D_{ka} h_j h_k + \sum_{k}  D_{ka}^2 h_k^2  \right]= \nonumber \\
& = \sum_a \sum_{k\ne i}  D_{ia}  D_{ka} h_i h_k + \frac{1}{2} \sum_a \sum_{k}  (D_{ia} D_{ka})^2 h_k^2 = \nonumber \\
& = h_i \sum_{k\ne i}  \Lambda_{ik} h_k +  \frac{1}{2}  \left [-  \sum_{k\ne i} \sum_a D_{ia} D_{ka} h_k^2 + \sum_a D_{ia}^2 h_i^2 \right ]  =\nonumber \\
& =  h_i \sum_{k\ne i}  \Lambda_{ik} h_k +  \frac{1}{2}  \left [ - \sum_{k\ne i}  \Lambda_{ik} h_k^2 + \Lambda_{ii} h_i^2\right ] =
\nonumber \\
& =  h_i \sum_{k}  \Lambda_{ik} h_k  -  \frac{1}{2}  \sum_{k}  \Lambda_{ik} h_k^2  \ ,
\end{align}
where we used the property (derived from \eqref{dprop}) $(D_{ia} D_{ka})^2 = - D_{ia} D_{ka} (1-\delta_{i,k}) + D_{ia}^2 \delta_{i,k}$ and the relation $\Lambda_{ij}=\sum_a D_{ia} D_{ja}$. From this we finally get
\begin{equation}
[(\nabla h)^2]_i = -  \frac{1}{2}  \sum_{k}  \Lambda_{ik} h_k^2  +  h_i \sum_{k}  \Lambda_{ik} h_k \ ,
\end{equation} 
which is Eq.~\eqref{grad2disc} of the main text. The definition of local scalar product given in this appendix is therefore fully consistent with the relationships between the gradient and the divergence operators in the continuum, exploited in Sec.~\ref{sec:grad2} to arrive at the same expression.

\section{The Random Euclidean graph}\label{app:A}

As discussed in the main text, a random Euclidean graph in $d=2$ is defined by the following procedure.
$N$ points are randomly and uniformly placed in a square of length $L$, giving a density of nodes $\rho=\frac{N}{L^2}$. The links between nodes are then generated according to a simple local rule: if two points are separated by a distance smaller than $r_c$ they are connected, otherwise they are not. In terms of the adjacency matrix, this reads
\begin{align}
\begin{cases}
   n_{ij}=1 \text{ }  \text{if } r_{ij} \leq r_c  \\
    n_{ij}=0 \text{ }  \text{if } r_{ij} > r_c  
    \end{cases}
\end{align}

Given a node in the graph, one can ask how many other sites are connected to it, i.e. what is the number $n$ of `interacting' neighbors (also called the `degree' of the node). Since points are uniformly drawn in space, the probability of this quantity is given by a binomial distribution, 
\begin{equation}
    p(n)=\binom{N}{n}r^n (1-r)^{N-n} \ ,
\end{equation}
where $r$ is the probability that two nodes are connected, i.e. $r=\dfrac{\pi r_c^2}{L^2}$ (when PBC are considered and $r_c < \dfrac{L}{2}$) \cite{dall2002random,penrose2003random}.  In the limit of large $N$ and low $r$, $p(n)$ tends to a Poisson distribution of the form
\begin{equation}
    p(n)=\frac{\langle n \rangle^n \exp{(-\langle n \rangle})}{n!} \ ,
\end{equation}
where $\langle n \rangle$ is the average number of interacting neighbors, given by $ \langle n \rangle = \rho \pi r_{c}^2$. 

In general, if the spatial density of points is too low (i.e. if the average nearest neighbor distance between points is much larger than the connectivity threshold $r_c$) the procedure described above might generate graphs that are divided into separate non-connected components. This is obviously not what we want. In our analyses we therefore considered values of $N$, $L$ and $r_c$ such that there is only one connected cluster (as we also verified numerically). In Fig.~\ref{fig:rand1} a) we show the distribution of interacting neighbors for two different sizes $N$ (one in which we are far from the Poisson limit and one in which we are in the Poisson limit) for density $\rho=4$ and $r_c=1$, i.e. the values used in the main text. The average number of interacting neighbors is $\langle n \rangle = \rho \pi r_{c}^2=12.56$.

Due to the intrinsic irregular distribution of points in space, there is no fixed lattice spacing. One can however consider as reference microscopic lenght-scale  the mean distance between closest neighbors $\langle d_{nn} \rangle$. This quantity scales with the density as $\rho^{-\frac{1}{2}}$ \cite{dall2002random,penrose2003random}, as illustrated in Fig. \ref{fig:rand1} b).

\end{document}